# Review Article: Integral Role of Physics in Advancing Pharmacy Education and Research

İzzet Sakallı*

1 AS245 Department of Physics, Eastern Mediterranean University, 99628, Famagusta Northern Cyprus, via Mersin 10, Turkiye.

## Abstract

Physics plays a fundamental role in advancing pharmacy education and research, providing theoretical underpinnings and practical tools necessary to address complex challenges in drug development, delivery, and diagnostics. This review explores the integration of physics into the pharmacy curriculum, highlighting how principles such as fluid dynamics, thermodynamics, and spectroscopy (Tokgöz and Sakallı, 2018) enhance students' critical thinking and problem-solving skills. Additionally, it examines the pivotal contributions of physics to pharmaceutical research, including molecular modeling, imaging technologies like MRI and PET, and nanotechnology-driven drug delivery systems. Despite challenges in interdisciplinary collaboration and resource allocation, innovative teaching strategies and laboratory-based learning are shown significant promise. Looking forward, the convergence of artificial intelligence and physics, as highlighted by recent Nobel Prize achievements in attosecond physics and bioorthogonal chemistry, is set to revolutionize pharmaceutical sciences, offering unprecedented precision and efficiency in drug discovery and personalized medicine.

### Keywords

## INTRODUCTION

Physics serves as a universal language for understanding and describing the fundamental laws of nature. Its principles permeate nearly every scientific discipline, offering a critical lens through which we explore complex phenomena. In the context of pharmacy, physics provides a foundational framework for numerous processes, ranging from drug formulation to delivery mechanisms and diagnostic technologies (Dhina et al., 2023; McCall, 2007). Despite its fundamental role, the explicit integration of physics into pharmacy education is often limited, with greater emphasis placed on chemistry and biological sciences. This oversight can obscure the value of physics as a key contributor to pharmaceutical advancements (Erdogan et al., 2021).

The historical evolution of pharmacy education has largely centered on the chemical and biological aspects of drug development, with physics playing a secondary role (Pillai and Cummings, 2013). However, as modern healthcare becomes increasingly reliant on interdisciplinary approaches, the importance of physics has grown exponentially. Drug design, for instance, is deeply rooted in physical chemistry, encompassing the study of molecular interactions, thermodynamics, and kinetics (Wilkinson et al., 2004). Similarly, the development of innovative drug delivery systems—such as nanoparticles and liposomes—requires an intricate understanding of mechanical forces, surface tension, and fluid dynamics (Blanke and Blanke, 1984).

Beyond its applications in drug formulation and delivery, physics also underpins critical diagnostic technologies that have revolutionized modern medicine. Techniques such as magnetic resonance imaging (MRI), computed tomography (CT), and positron emission tomography (PET) rely on principles of nuclear physics and electromagnetism (McCall, 2007; Gulcan et al., 2021). These technologies not only aid in early disease detection but also enhance our understanding of physiological processes, paving the way for more targeted and effective treatments (Pepeu and Giovannini, 2009).

From a pedagogical perspective, integrating physics into pharmacy education can significantly enhance students' problem-solving and analytical skills (Dhina et al., 2023). The study of physics trains individuals to approach problems systematically, breaking down complex systems into comprehensible components (Erdogan et al., 2021). For pharmacy students, this skillset is invaluable in navigating the multifaceted challenges of drug development, quality control, and clinical application (Gulcan et al., 2019). However, achieving this integration requires a paradigm shift in how physics is taught within the pharmacy curriculum (Wilkinson et al., 2004). Traditional lecture-based methods must be supplemented with context-driven approaches that highlight the relevance of physics to pharmaceutical sciences (Shukur et al., 2021).

Several studies have highlighted the potential benefits of such an integrated approach. For instance, the use of case studies that link physics concepts to real-world pharmaceutical problems has been shown to improve student engagement and comprehension (Ellman, 1958). Laboratory experiments, too, can serve as a powerful tool for demonstrating the practical applications of physics, from measuring drug solubility and diffusion rates to analyzing the mechanical properties of tablet formulations (Blanke and Blanke, 1984; McCall, 2007).

Despite these advantages, significant barriers remain. One of the primary challenges is the lack of interdisciplinary collaboration between physics and pharmacy departments (Gulcan et al., 2003). Bridging this gap requires concerted efforts to develop curricula that are both scientifically rigorous and practically relevant (Pepeu and Giovannini, 2009). Additionally, faculty development programs can play a crucial role in equipping educators with the skills needed to effectively teach physics in a pharmaceutical context (Pillai and Cummings, 2013; Wilkinson et al., 2004).

In conclusion, physics is not merely a supplementary discipline in pharmacy education but a cornerstone for understanding and advancing the pharmaceutical sciences (Erdogan et al., 2021). As the healthcare landscape continues to evolve, the integration of physics into pharmacy education will be essential for preparing the next generation of pharmacists to meet the demands of an increasingly complex and interdisciplinary field (Gulcan et al., 2021).

The review paper is organized as follows. Section 2 elaborates on how foundational physics concepts are integral to understanding drug formulation and delivery mechanisms, supported by examples and measurable learning outcomes. In section 3, the applications of physics in cutting-edge research, including molecular modeling, nanotechnology, and diagnostic imaging, are discussed with real-world implications. Section 4 highlights the need for innovative teaching methods, interdisciplinary collaboration, and laboratory-based learning to effectively incorporate physics into pharmacy education. The review concludes with section 5, which is a summary of the key findings, emphasizing the future directions for integrating artificial intelligence and recent advances in physics, such as attosecond technologies, to revolutionize pharmaceutical sciences.

## IMPORTANT OF PHYSICS IN PHARMACY EDUCATION

Physics serves as the backbone for numerous applications in pharmacy education, providing fundamental insights into the physical principles governing pharmaceutical systems. From understanding drug solubility to designing advanced diagnostic tools, the role of physics cannot be overstated. This section highlights the importance of physics in the pharmacy curriculum, discussing key concepts, formulas, and real-world examples that underline its relevance.

One of the most direct applications of physics in pharmacy is the study of fluid dynamics. Fluid flow is critical in understanding how drugs disperse in the human body, particularly through the circulatory and lymphatic systems (McCall, 2007; Wilkinson et al., 2004). For instance, Poiseuille's Law, which governs laminar flow, can be expressed as:

$$Q = \pi r^4 \Delta P / (8\mu L)$$

Here, $Q$ represents the flow rate, $r$ is the radius of the tube (e.g., blood vessel), $\Delta P$ is the pressure difference, $\mu$ is the viscosity, and $L$ is the length of the tube. This equation highlights how minor changes in the radius of blood vessels can significantly impact the flow rate, which is a critical consideration for intravenous drug delivery.

Thermodynamics, another essential branch of physics, provides theoretical framework for understanding drug stability and solubility. The Gibbs free energy ($\Delta G$) equation (Pourhassan et al., 2023), often used in pharmaceutical studies, is given as:

$$\Delta G = \Delta H - T\Delta S$$

This equation helps predict whether a reaction or process will occur spontaneously (Dhina et al., 2023; Erdogan et al., 2021). For example, the solubility of a drug in a solvent depends on the interplay between enthalpy ($\Delta H$) and entropy ($\Delta S$) changes, influencing formulation strategies and storage conditions.

Physics also plays a pivotal role in the design and optimization of medical imaging technologies, such as magnetic resonance imaging (MRI) and computed tomography (CT) (Blanke and Blanke, 1984; Gulcan et al., 2021). MRI, for instance, relies on principles of nuclear magnetic resonance. The Larmor frequency, which describes the procession of nuclear spins in a magnetic field, is expressed as:

$$\omega = \gamma B$$

Here, $\omega$ is the angular frequency, $\gamma$ is the gyromagnetic ratio, and $B$ is the magnetic field strength.

Spectroscopic techniques such as ultraviolet-visible (UV-Vis) and infrared (IR) spectroscopy are foundational in pharmaceutical analysis. The Beer-Lambert law is widely used to determine the concentration of drug solutions and is expressed as:

$$A = \varepsilon c l$$

In this equation, $A$ represents absorbance, $\varepsilon$ is the molar absorptivity, $c$ is the concentration of the solution, and $l$ is the path length of the light through the sample. This law is critical for ensuring accurate dosage in liquid formulations.

Rheology, the study of flow and deformation of matter, is integral to the development of various pharmaceutical dosage forms such as creams, ointments, and gels (Pepeu and Giovannini, 2009; McCall, 2007). Understanding the shear-thinning or shear-thickening behavior of these formulations ensures their stability and efficacy. For instance, shear-thinning behavior, where viscosity decreases with increasing shear rate, is critical for injectable formulations, enabling ease of administration through syringes.

To summarize, physics is a cornerstone of pharmacy education, providing the theoretical and practical tools necessary to address complex pharmaceutical problems. By integrating physics concepts into the curriculum, educators can empower future pharmacists to innovate and excel in their field (Blanke and Blanke, 1984; Erdogan et al., 2021).

## ROLE OF PHYSICS IN ADVANCED PHARMACEUTICAL RESEARCH AND TECHNOLOGY

Physics plays a transformative role in pharmaceutical research, bridging the gap between theoretical principles and practical applications. The incorporation of physics into research has enabled significant advancements in drug discovery, formulation, delivery mechanisms, and diagnostic technologies. This section explores the pivotal role of physics in pharmaceutical research, highlighting key concepts, equations, and real-world applications.

One of the foundational applications of physics in pharmaceutical research is the study of molecular interactions using quantum mechanics. Techniques such as molecular docking and simulations rely on the Schrödinger equation to predict the behavior of molecules at the atomic level. The time-independent form of the Schrödinger equation is given as:

$$H\Psi = E\Psi$$

Here, $H$ is the Hamiltonian operator representing the total energy of the system, $\Psi$ is the wave function, and $E$ is the energy eigenvalue. By solving this equation, researchers can predict molecular conformations and interactions, guiding the development of more effective drug candidates (Erdogan et al., 2021; Gulcan et al., 2021).

Thermodynamics also plays a crucial role in understanding drug stability and solubility. The van 't Hoff equation, used to analyze the temperature dependence of equilibrium constants, is expressed as:

$$ln(K) = -\Delta H/RT + C$$

In this equation, $K$ is the equilibrium constant, $\Delta H$ is the enthalpy change, $R$ is the gas constant, $T$ is the temperature, and $C$ is a constant. Understanding this relationship allows researchers to optimize conditions for drug formulations, ensuring their stability and efficacy (Wilkinson et al., 2004; Blanke and Blanke, 1984).

The study of fluid dynamics is indispensable in designing drug delivery systems. The Navier-Stokes equations describe the motion of viscous fluids, which is critical for understanding blood flow and drug dispersion. A simplified version of these equations for incompressible flow is:

$$\rho(\partial v/\partial t + v \cdot \nabla v) = -\nabla p + \mu\nabla^2 v$$

Here, $\rho$ is the fluid density, $v$ is the velocity vector, $p$ is the pressure, and $\mu$ is the dynamic viscosity. These equations are used to model the behavior of injectable drugs and predict their distribution within the body (McCall, 2007; Dhina et al., 2023).

Physics also contributes to nanotechnology-based drug delivery systems. The application of optical tweezers, which relies on the principles of electromagnetic radiation pressure, has facilitated the manipulation of nanoparticles for targeted drug delivery. The trapping force in optical tweezers is given by:

$$F = (nP/c)(1 + R - T)$$

In this equation, $F$ is the force, $n$ is the refractive index, $P$ is the power of the laser beam, $c$ is the speed of light, $R$ is the reflectivity, and $T$ is the transmissivity. Such technologies enhance precision in delivering drugs to specific sites, minimizing side effects (Gulcan et al., 2019; Ellman, 1958).

Imaging technologies have also benefited from advancements in physics. Positron Emission Tomography (PET) and Magnetic Resonance Imaging (MRI) are vital tools in pharmaceutical research, enabling non-invasive monitoring of drug behavior within the body. The signal-to-no ise ratio (SNR) in MRI, a critical parameter for image quality, is influenced by the following relationship:

$$SNR \propto B_0^2 \sqrt{(\Delta V)}$$

Here, $B_0$ is the magnetic field strength, and $\Delta V$ is the voxel volume. Higher $B_0$ fields improve image resolution, allowing detailed observation of drug interactions and effects (Pepeu and Giovannini, 2009; Shukur et al., 2021).

To summarize, physics underpins a wide range of innovations in pharmaceutical research, from molecular modeling to advanced imaging techniques. Its principles provide the tools to address complex challenges, paving the way for more effective drugs and delivery systems (Erdogan et al., 2021; McCall, 2007).

## INTEGRATION OF PHYSICS INTO PHARMACY CURRICULUM

Integrating physics into the pharmacy curriculum is essential to equipping students with the foundational knowledge and critical thinking skills required to address complex challenges in pharmaceutical sciences. By understanding key physical principles, pharmacy students gain insights into drug formulation, delivery mechanisms, and diagnostic technologies. This section explores the strategies and outcomes associated with integrating physics into the pharmacy curriculum, along with examples of learning outcomes and real-world applications.

Physics courses tailored for pharmacy students should emphasize the relevance of physical principles to pharmaceutical applications. For instance, understanding thermodynamics can help students predict drug stability under various conditions, while knowledge of fluid dynamics can aid in modeling blood flow and drug dispersion. These concepts are critical in both clinical and research settings (McCall, 2007; Erdogan et al., 2021).

A well-structured curriculum should also focus on measurable learning outcomes. Examples of physics-related learning outcomes for pharmacy students include:

1. Demonstrating an understanding of the principles of fluid dynamics and their

application to blood flow and intravenous drug delivery.

2. Applying thermodynamic principles to evaluate drug solubility, stability, and shelf life.

3. Utilizing spectroscopic techniques, such as UV-Vis and IR spectroscopy, to analyze pharmaceutical formulations.

4. Explaining the role of nuclear physics in medical imaging technologies, including MRI and PET scans.

5. Modeling molecular interactions using quantum mechanics to predict drug efficacy and receptor binding.

Innovative teaching methods, such as problem-based learning (PBL) and case studies, can enhance the integration of physics into the pharmacy curriculum. For instance, PBL activities could involve analyzing the diffusion of a drug through a semipermeable membrane or designing a nanoparticle-based drug delivery system (Shukur et al., 2021; Gulcan et al., 2021). Such approaches encourage active learning and help students connect theoretical knowledge to practical applications.

The use of laboratory experiments is another effective way to teach physics in the context of pharmacy. Examples of experiments include measuring the diffusion coefficients of drug molecules, analyzing fluid viscosity, and investigating the thermal properties of pharmaceutical materials. These hands-on experiences not only reinforce theoretical concepts but also prepare students for real-world challenges in research and clinical practice (Dhina et al., 2023; Wilkinson et al., 2004).

Assessment strategies should also be aligned with the integration of physics into the pharmacy curriculum. Traditional exams can be supplemented with project-based assessments that require students to solve complex, interdisciplinary problems. For example, a project could involve designing a drug delivery system that considers fluid dynamics, thermodynamics, and material properties (Blanke and Blanke, 1984; Pepeu and Giovannini, 2009).

Despite its importance, the integration of physics into the pharmacy curriculum faces several challenges. These include a lack of interdisciplinary collaboration between departments, limited faculty expertise in physics, and insufficient resources for laboratory-based instruction. Addressing these challenges requires institutional support, faculty development programs, and investments in state-of-the-art teaching facilities (Ellman, 1958; Pal et al., 2023).

In conclusion, integrating physics into the pharmacy curriculum is vital for developing the analytical and problem-solving skills necessary for success in pharmaceutical sciences. By adopting innovative teaching strategies and aligning learning outcomes with industry needs, educators can prepare the next generation of pharmacists to excel in both academic and clinical settings (Erdogan et al., 2021; McCall, 2007).

## CONCLUSION

This review has explored the multifaceted role of physics in pharmacy education and research, emphasizing its foundational importance in understanding and advancing pharmaceutical sciences. Physics provided the theoretical and practical frameworks necessary for addressing challenges in drug design, delivery, diagnostics, and education. In pharmacy education, physics was integrated to equip students with critical thinking skills and interdisciplinary knowledge. By learning key principles such as fluid dynamics, thermodynamics, and spectroscopy, students gained a comprehensive understanding of drug stability, solubility, and delivery mechanisms. Laboratory experiments and problem-based learning further reinforced these principles, connecting theory to practical applications.

In pharmaceutical research, the application of physics was shown to bridge the gap between theoretical models and clinical outcomes. From the Schrödinger equation guiding molecular docking to the Navier-Stokes equations modeling fluid flow, physics was instrumental in innovating drug delivery systems, imaging technologies, and nanotechnology-based approaches. Imaging modalities such as MRI and PET relied on physical principles, enhancing the precision of diagnostics and treatment monitoring. Furthermore, techniques like optical tweezers and spectroscopy enabled targeted drug delivery and analysis, minimizing side effects and optimizing therapeutic efficacy.

Integrating physics into the pharmacy curriculum faced challenges, including the need for interdisciplinary collaboration and resources. Despite these hurdles, the adoption of innovative teaching strategies, such as problem-based learning and laboratory-based instruction, was demonstrated to significantly enhance the educational experience. These approaches not only prepared students for research and clinical practice but also cultivated a deeper appreciation of the relevance of physics in their professional roles.

Looking ahead, the integration of artificial intelligence (AI) into the intersection of physics and pharmacy presents transformative potential. AI algorithms, particularly those leveraging advancements in quantum computing, are expected to revolutionize molecular modeling, drug discovery, and personalized medicine. The 2024 Nobel Prize in Physics (Hopfield & Hinton, 2024) highlighted breakthroughs in attosecond physics, which promise to refine imaging techniques and drug-target interactions at unprecedented temporal resolutions. Similarly, the Nobel Prize in Chemistry (Baker et al., 2024) celebrated advancements in bioorthogonal chemistry, which, when coupled with AI-driven insights, could lead to more precise and dynamic pharmaceutical interventions.

The future of this field will likely involve a deeper integration of AI and physics, creating new paradigms in pharmaceutical sciences. By harnessing the computational power of AI, researchers can accelerate the development of drugs and diagnostics, optimize delivery mechanisms, and predict patient-specific outcomes with greater accuracy. This confluence of disciplines will not only enhance the effectiveness of therapeutic interventions but also redefine the educational frameworks that prepare future pharmacists for this rapidly evolving landscape.

## DECLARATION

I, İzzet Sakallı, hereby declare that this manuscript is my original review article and has not been published or submitted elsewhere in any form. I confirm that all sources used in the preparation of this article have been appropriately cited and referenced. There is no conflict of interest related to the publication of this work. I take full responsibility for the accuracy of the content and conclusions presented in this manuscript. Additionally, I acknowledge that the research and analysis presented in this paper align with ethical standards and academic integrity.

# REFERENCES

Baker, D., Hassabis, D., & Jumper, J. (2024). *The Nobel Prize in Physics 2024. Retrieved from* https://www.nobelprize.org/prizes/physics/2024/summary/.

Blanke, S. R., & Blanke, R. V. (1984). The Schotten-Baumann reaction as an aid to the analysis of polar compounds: application to the determination of tris (hydroxymethyl) aminomethane (THAM). *Journal of Analytical Toxicology, 8(5), 231–233*.

Dhina, M. A., Kaniawati, I., & Yustiana, Y. R. (2023). Learning basic physics in Pharmacy Study Program with systems thinking skills needed by pharmacy students. *Momentum: Physics Education Journal, 8(1), 55–64*.

Ellman, G. L. (1958). A colorimetric method for determining low concentrations of mercaptans. *Archives of Biochemistry and Biophysics, 74(2), 443–450*.

Erdogan, M., Kilic, B., Sagkan, R. I., Aksakal, F., Ercetin, T., et al. (2021). Design, synthesis and biological evaluation of new benzoxazolone/benzothiazolone derivatives as multi-target agents against Alzheimer's disease. *European Journal of Medicinal Chemistry, 212, 113124*.

Gulcan, H. O., & Orhan, I. E. (2021). Dual Monoamine Oxidase and Cholinesterase Inhibitors with Different Heterocyclic Scaffolds. *Current Topics in Medicinal Chemistry, 21(30), 2752–2765*.

Gulcan, H. O., Mavideniz, A., Sahin, M. F., & Orhan, I. E. (2019). Benzimidazole-derived compounds designed for different targets of Alzheimer's disease. *Current Medicinal Chemistry, 26(18), 3260–3278*.

Hopfield, J. J., & Hinton, G. (2024). *The Nobel Prize in Physics 2024. Retrieved from* https://www.nobelprize.org/prizes/physics/2024/summary/.

McCall, R. P. (2007). Relevance of physics to the pharmacy major. *American Journal of Pharmaceutical Education, 71(4), Article 70*.

Pal, R., Pandey, P., & Amjad, T. M. (2023). The dominance role of physics in pharmaceutical dosage form formulations. *Goya Journal, 16(5), 125–138*.

Pillai, J. A., & Cummings, J. L. (2013). Clinical trials in predementia stages of Alzheimer disease. *Medical Clinics, 97(3), 439–457*.

Pourhassan, B., Hendi, S. H., Upadhyay, S., Sakalli, I., & Saridakis, E. N. (2023). Thermal fluctuations of (non)linearly charged BTZ black hole in massive gravity. *Int. Jour. Mod. Phys. D, 32 (16), 2350110*.

Shukur, K., Ercetin, T., Luise, T., Sippl, C., Sirkecioglu, W., et al. (2021). Design, synthesis, and biological evaluation of new urolithin amides as multitarget agents against Alzheimer's disease. *Archiv der Pharmazie, 354(5), 2000467*.

Tokgoz, G., & Sakalli, I. (2018). Spectroscopy of *z=0* Lifshitz Black Hole. *Adv. High Energy Phys., 2018*, 3270790

Wilkinson, D. G., Francis, P. T., Schwam, E., & Payne-Parrish, J. (2004). Cholinesterase inhibitors used in the treatment of Alzheimer's disease. *Drugs & Aging, 21(7), 453–478*.